# Conductivity crossover in nano-crystalline diamond films: Realization of a disordered superlattice-like structure.

GEORGE CHIMOWA, DIMITRY CHUROCHKIN AND SOMNATH BHATTACHARYYA[(A)]

*Nano-Scale Transport Physics Laboratory, School of Physics and DST/NRF Centre of Excellence in Strong Materials, University of the Witwatersrand, Private Bag 3, WITS 2050, Johannesburg, South Africa*

**Abstract** – We present the electrical transport characteristics of a batch of nano-crystalline diamond films of varying nitrogen concentrations and explain the conduction mechanism by the disordered quasi-superlattice model applied to semiconductor heterostructures. Synthesized by the hot filament chemical vapour deposition technique, the degree of structural disorder in the films, confirmed from Raman spectroscopy, is found to be controllable, resulting in the transition of conduction mechanism from localized and activated to the metallic conduction regime. Hence through high field magneto-resistance measurements at low temperatures we firmly establish a conductivity crossover from hopping to 3D weak localization. The long electronic dephasing time and its weak temperature dependence suggest the possibility for diamond-based high-speed device applications.

**Introduction.** – Nitrogen-doped ultranano-crystalline diamond (UNCD) films prepared by microwave plasma enhanced chemical vapor deposition (MPCVD) have been studied extensively and found to be very promising in nano-electronic device applications [1-2]. Generally, the conductivity of these films can be controlled to some extent with the increase of nitrogen level [3] however, it has appeared to be very different from conventional semiconductors. Further studies are therefore required in order to establish the (nitrogen) doping mechanism in these films [4]. Previous studies by researchers on heavily nitrogen doped UNCD films prepared by MPCVD showed metallic behavior along with negative magneto-resistance (MR) features explained using the 2D weak localization (WL) or hopping model [1,2,5]. In this article we report activated conduction over a wide range of temperatures which has not been demonstrated previously, but it is now observed in NCD samples grown by hot filament chemical vapor deposition (HFCVD) technique.

Inclusion of amorphous carbon in nanodiaomond films increases the density of localized states in the band gap and makes the hopping processes dominant in these materials [6]. This observation clearly emphasizes the importance of the $sp^2$ phases in the grain boundaries (GB) [6], which depend significantly on the gas chemistry during synthesis. Furthermore, detailed microscopic studies in UNCD films by other researchers have suggested superlattice-like structure consisting of layers of $sp^2$ bonded carbon within the GB separated by diamond grains [7,8]. We extended the idea of previous researchers and employed a disordered quasi-superlattice model that was applied previously to semiconductor heterostructures, which is discussed later. While a great deal of attention has been given to synthesis, microstructure and morphology studies, little has been reported on the low temperature transport properties of nitrogen-doped NCD films grown particularly by HFCVD [9]. In this work we therefore attempt to explain the electrical transport properties of these HFCVD films (whose grain boundary is different from MPCVD films because of the different gas chemistry) and establish the validity of the recently reported anisotropic 3D WL model in nitrogen doped nano-crystalline diamond films synthesized by methods other than MPCVD [10]. We report a transition from hopping to activated conduction and finally an anisotropic 3D WL mechanism in the transport properties of the films, as the nitrogen percentage in the HFCVD chamber is increased. With the help of a disordered superlattice model and microstructure studies using Raman spectroscopy we show how disorder controls the conductivity and the characteristic time of NCD films.

**Experimental Techniques.** – The nitrogen incorporated nano-crystalline diamond films were synthesized in a commercial HFCVD chamber (Vacutec, SA) at a substrate temperature of approximately 800 °C by introducing 10% to 22% ultra-pure nitrogen (99.999%) in the reaction zone, which are labeled $NCD10N_2$, $NCD15N_2$, $NCD20N_2$ and

[(A)] E-mail: Somnath.Bhattacharyya@wits.ac.za

NCD22N$_2$. The other reaction gases were methane kept at (4.5%), hydrogen and variable argone concentrations. The argone flow was varied according to the nitrogen flow rate so as to keep the total volume flow at 210 sccm. The low pressure (22 mBar) and relatively high methane concentration are conducive to the formation of nanoscale grains, possibly due to the high renucleation rate [9]. Four probe conductivity-temperature measurements of these samples (0.4 cm × 0.3 cm, thickness ~ 0.165 μm) deposited on fused quartz substrates were performed in the temperature range from 2.3 K to 300 K and applying a magnetic field up to 12 T using a fully automated cryogenic free measurement system. A current of 10 μA was sourced from a Keithly 2400 and the voltage was measured using a Keithly 2182A nano-voltmeter using the van der Pauw configuration. Raman spectroscopy was performed using an Argon ion laser (514.5 nm) at an average power of ~1 mW. Details of NCD film preparation and microscopic characterization are given elsewhere [11].

**Results and Discussion.**–

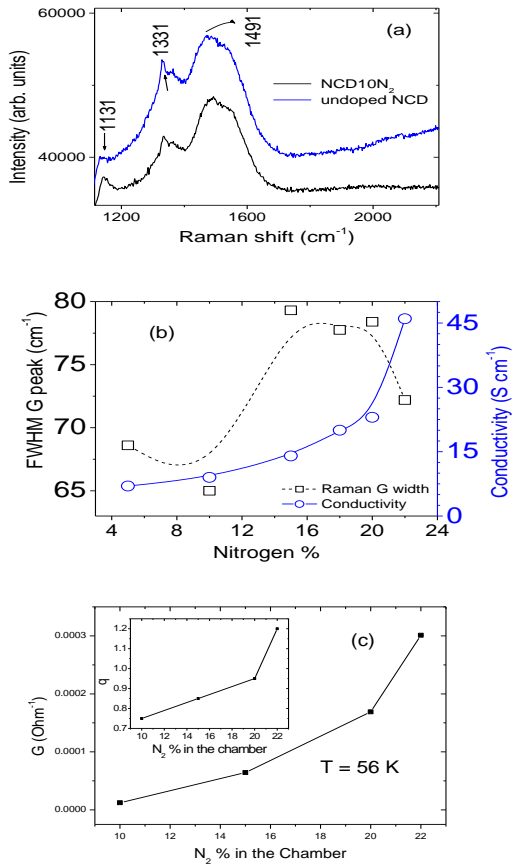

Fig. 1: **(a)** Raman spectra for doped and undopednanodiamond films prepared by HFCVD technique. **(b)** Variation of G linewidth with nitrogen concentration in the film deposition chamber **(c)** Conductance dependence on N$_2$% in the chamber shows a continuous increase. The insert shows the variation of the disorder parameter $q$ with N$_2$%.

It is evident from the Raman spectra that nitrogen introduction in the chamber significantly reduces the intensity of the diamond peak at (1332 cm$^{-1}$), which is an indication of increase of the width of grain boundary regions in the films. Figure 1(b) also shows the change of Raman G peak width with N$_2$% used for synthesis of NCD films, which is an estimation of disorder level in the films **[12].** Although disorder initially increases with the N$_2$%, it decreases rapidly in the high N$_2$% regime as observed in Raman spectra of the samples. Conductance of the samples shows an increase with N$_2$% in Fig. 1(c). The corresponding change of disorder parameter ($q$) is explained from the analysis of conductance data in the later part of the text. Transmission electron micrographs indicated that the grain sizes are in the range from 15 to 20 nm [11].

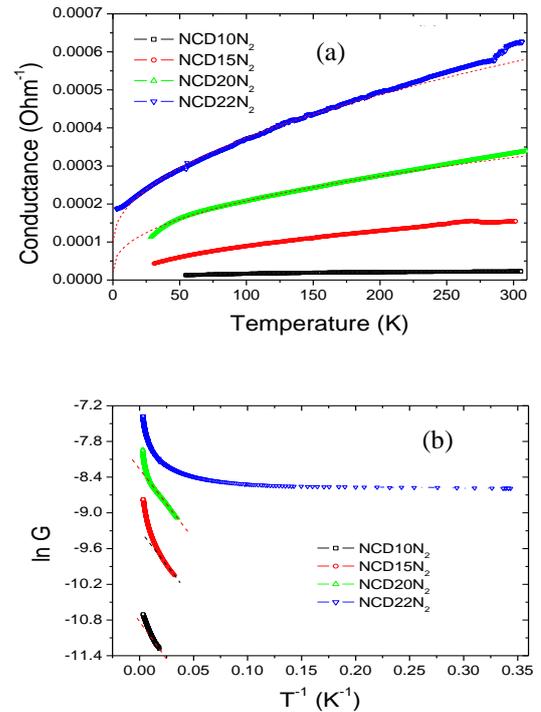

Fig. 2: **(a)** Conductance *vs.T* fitted with 3D WL model for NCD films (dotted lines). **(b)** Arrhenius plot for G(*T*) to validate activated conduction.

**Electrical transport:** Figures 2(a) and 2(b) show three distinct behaviors at low temperatures namely, weakly conducting (for NCD10N$_2$ and NCD15N$_2$), semiconducting (NCD20N$_2$) and semi-metallic (NCD22N$_2$) films, respectively. To understand the transport mechanism, plots of ln *R* vs. $T^{\chi}$ were done (not shown here), with χ being 1/2, 1/3 or 1/4 for 1D, 2D or 3D hopping respectively. The analysis showed 3D VRH hopping is applicapable for both NCD10N$_2$ and NCD15N$_2$ samples below 118 K. While Arrhenius plots for all the films in Fig. 2(b) clearly show evidence of activated conduction from 28 K to 89 K for the NCD20N$_2$ films.

This thermally activated conduction is normally expressed in the form of Arrhenius equation as $\sigma(0,T) = \sigma_o exp(-\frac{\Delta E}{kT})$, where $\Delta E$ is the conductivity activation energy and $\sigma_o$ is the conductivity prefactor [13]. The activation energy calculated from Fig. 2(b) is found to be less than 10 meV. The origin of the energy gap can be explained from the effect of disorder on the nitrogen donor level formed close to the conduction band. The hopping mechanism and activated behaviour were not found to be applicable for $NCD22N_2$ films. In earlier reports, N-UNCD films were treated as disordered metals or bulk amophous semiconductors where confinement properties in GB were not included [4,5]. Here, we interpret the transport in the family of NCD films of different conducitivty using a quasi-SL structure. Fig. 2(a) shows the conductance *vs.* temperature data fitted with Eq.(1). In the conventional 3D WL isotropic model the temperature dependence of the conductance ($G$) is expressed as a sum of the temperature independent term ($G_0$) as well as the terms consisting of the temperature dependence of the dephasing length and 3D electron-electron (*e-e*) interaction given by

$$G(0,T) = G_o + a_1 T^{0.35} + a_2 T^{0.5} \qquad (1),$$

where, $a_1 = \frac{S}{l}\left(e^2/2\pi^2\hbar L_\phi\right)$ and $a_2 = \frac{S}{l}\left(\frac{e^2}{4\hbar\pi^2}\right)\left(\frac{1.3}{\sqrt{2}}\right)\left(\frac{4}{3} - \frac{3}{2}F\right)\sqrt{(k_B/\hbar D)}$, are pre-factors that depend on sample dimensions given on page 2 ($S$, cross-sectional area and $l$, length), dephasing length ($L_\phi$) and diffusion constant ($D$) [**14**]. The terms $F, \tau,$ and $D$ represent electron screening factor in 3D, the relaxation time for the *e-e* interactions, and the diffusion constant, respectively [**14**]. From the fitting of *G-T* data [Fig. 2(a)] the values of $a_1$ and $a_2$ were evaluated from which $L_\phi$ can be derived as ~ $1.12 \times 10^{-8} T^{-0.33}$ (m) for $NCD20N_2$ and ~ $3.89 \times 10^{-8} T^{-0.35}$ (m) for $NCD22N_2$ films. Furthermore the weak temperature dependence of the dephasing factor compared to the *e-e* interaction in equation (1) shows that in the semi-metallic regime, *e-e* interactions dominate over the quantum interference. These observations are supported by theoretical work by other reseachers who predicted the contribution of the nitrogen related centres to the conductivity by determining the electronic stuctures of several nitrogen centres (defect centres) in nanodiamond films [**6,7**]. Here we present a detailed explanation of the conduction in n-NCD films, in the context of a disorderd superlattice like structure.

**Magneto-resistance:** MR features in Fig 3(a) and (b) for $NCD10N_2$ and $NCD15N_2$, respectively, are found to be very weakly *B* dependent. At high *T* the MR is found to be linear with *B*, which is consistent with the 3D VRH mechanism. This confirms the applicability of the VRH mechanism suggested earlier for the weakly conducting samples. The MR data were fitted with Eq. (2),

$$ln\left(\frac{\rho_B}{\rho_0}\right) = -a_1 B + a_2 B^2 + a_3 \qquad (2),$$

and using the constant $T_0$, the localization length was obtained for $NCD10N_2$ samples to be 11.98 nm at 100 K. The parameters $a_1 = NL_c{}^5\left(\frac{T_0}{T}\right)^{P1}\left(e/\hbar\right)$, $a_2 = \left(\frac{5e^2 L_c{}^4}{2016\hbar^2}\right)\left(\frac{T_o}{T}\right)^{P2}$ and $a_3$ account for the complex behavior of MR at *B*. The indexes *P1* and *P2* are 7/8 and 3/4, respectively in the Mott VRH model [**15**]. These results are consistent with a previous report on MPCVD-UNCD samples prepared with approximately 1% nitrogen [**4**]. At present we have observed a transition from VRH to activated conduction with increase in nitrogen percentage, which suggests band modification in these films as a result of nitrogen incorporation and associated structural rearrangement of carbon atoms in the grain boundaries.

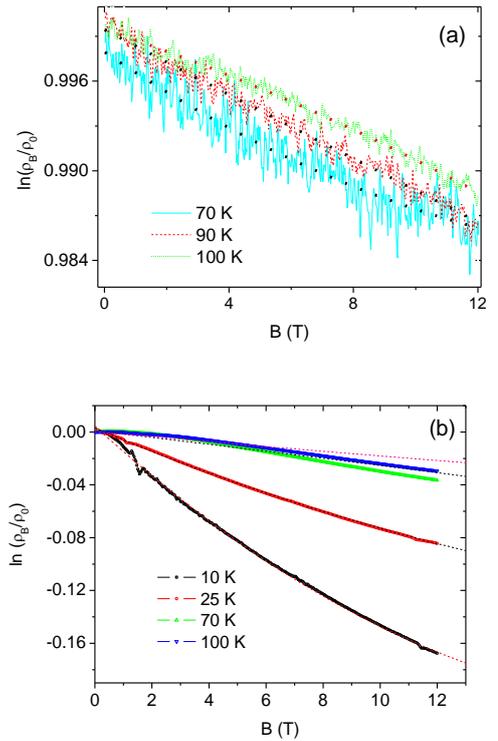

Fig.3:**(a)** ln $(\rho_B/\rho_0)$ *vs.B* graph for $NCD10N_2$ films and **(b)** $NCD15N_2$ films fitted with Eq. (2) to evaluate the localization lengths.

The MR data (Fig.3(b)) for $NCD15N_2$ samples show a clear deviation from the VRH fit at temperatures above 50 K. We estimated the $N(E_F)$ ~ $2.54 \times 10^{19}$ eV$^{-1}$ cm$^{-3}$ at 100 K with the value of $L_c$ ~ 11.9 nm, which is slightly greater than $6.46 \times 10^{18}$ eV$^{-1}$ cm$^{-3}$ obtained for $NCD10N_2$ samples. These results are consistent with the observation of increase of conductivity in the films prepared with higher $N_2$% [see Fig. 1(b)].
Analysis of the $NCD20N_2$ MR data showed a $B^{1/2}$ dependence of the MR at high fields, which is characteristic of 3D WL [**16,17**]. This is in agreement with the reports by other researchers that have indicated a $B^{1/2}$ or $B^2$ dependence of MR for UNCD films [10]. The magnitude of this MR decreases with an increase in temperature indicating that WL effects are suppressed at high temperatures in these films. We further attempt to validate the recently reported 3D WL anisotropic model used to explain the conductivity in nano-diamond films [10]. This behavior has been explained in terms of a propagative fermi surface (PFS).

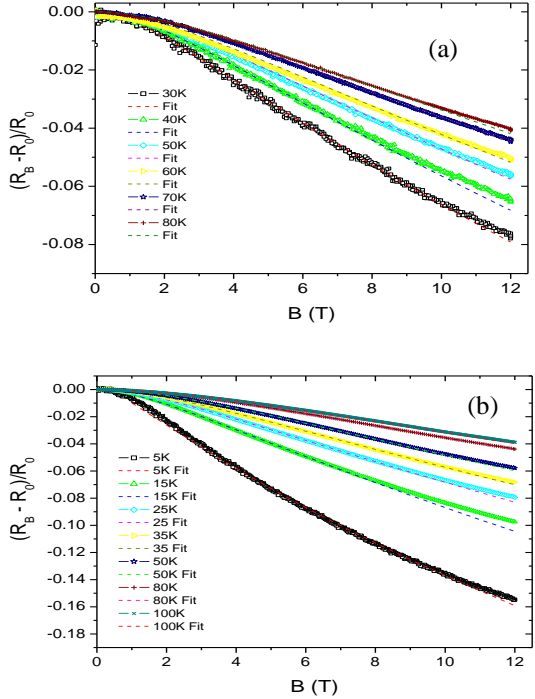

Fig. 4: MR results with 3D WL anisotropic fitting for (a) NCD20N$_2$ and (b) NCD22N$_2$ films.

The PFS model was originally developed to explain transport in disordered artificial SL and was recently used to explain the unusual transport in nano-crystalline silicon films in the diffusive fermi surface (DFS) limit considering an effective artificially formed SL structure [18]. The behavior is given by:

$$\Delta\sigma(B)/\sigma(0,T) = \alpha\left[\sigma(0,T)\left(\frac{e^2}{2\hbar\pi^2}\right)\left(\frac{eB}{\hbar}\right)^{1/2}\right]f_3\left\{\frac{\hbar/eB}{4D\tau_\phi}\right\} \quad (3),$$

The function $f_3$ is the Kawabata function [16]. The anisotropy coefficient $\alpha = \sqrt{D_{||}/D_\perp}$ describes anisotropic transport in 3D, where $D_{||}$ and $D_\perp$ represent the diffusion coefficient parallel and perpendicular to the film, respectively. This function predicts a $B^2$ dependence at low fields and a $B^{1/2}$ at high fields. Normalized MR data fitted with Eq. (3) confirms that the 3D WL anisotropic model best describes the MR ($B,T$) behavior for NCD20N$_2$ and NCD22N$_2$ films see [Fig 4(a) & (b)]. The anisotropic coefficient for the films was found to be 1.5 for the NCD20N$_2$ and 2.1 for the NCD22N$_2$ films. Our analysis indicates that the anisotropy factor is less in weakly conducting films than that for highly conducting films in agreement with previous reports [18]. We explain this from the structural arrangement induced by nitrogen doping leading to the appearance of the layered structure in NCD films at high levels of N$_2$ concentration [8]. Consequently the highly conducting NCD films prepared with a high N$_2$ percentage can consist of a large number of layers and show a strong anisotropic nature. From the fitting of the MR data, $L_\phi$ can be extracted whose temperature dependence i.e. $L_\phi \sim T^{-0.33}$ is plotted [Fig. 5(a)] which is similar to that obtained from $G(T)$. These results confirm the applicability of the anisotropic 3D WL model to explain the conduction mechanism of highly conducting NCD-N$_2$ films in the low to intermediate temperature range. Now we attempt to show the effect of disorder on the proposed quasi-SL model and investigate the origin of the observed activated conduction in NCD20N$_2$ films, which appeared as a result of nitrogen incorporation. The band diagram is typical of a disordered superlattice-like structure formed by nano-diamond grains (sp$^3$) and sp$^2$ carbon layers mainly in the GB. Transport calculations on N-doped disordered carbon SL structures are presented in Ref. **10**. Increase in nitrogen in the films results in widening of GB, which consequently increases the sp$^2$ layers and hence an incresease in the coupling of the diamond grains. In addition to that some nitrogen defects states are also introduced which might explain the activated behaviour in some samples [Fig. 5(b)] [**3,7**].

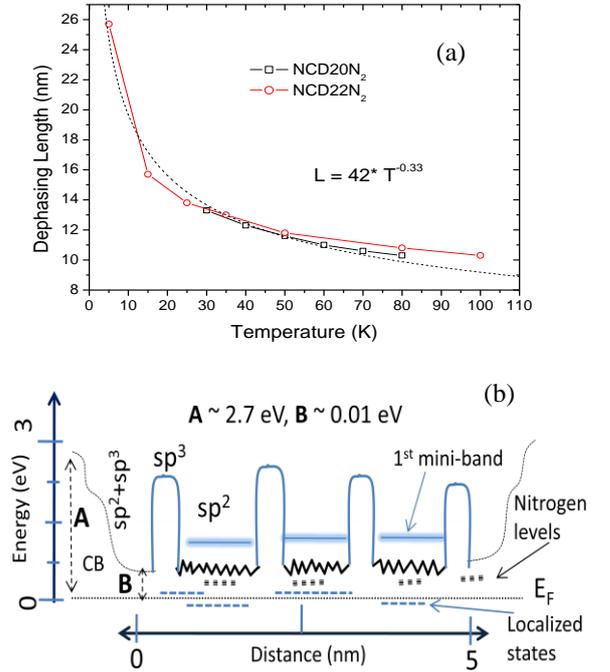

Fig. 5 **(a)**: The temperature dependence of the dephasing length for NCD20N$_2$ and NCD22N$_2$ samples i.e. $L_\phi \sim T^{-0.33}$ (see dotted lines). **(b)**: Band energy diagram of NCD films shows disordered sp$^2$ layers separated by local (sp$^3$) barriers within the GB between two diamond grains. The typical energy levels of localized states and nitrogen levels are marked in the diagram.

It is well known that nitrogen incorporation in carbon films introduces sp$^2$-bonded clusters [**8**]. The microstructure of NCD films show confined sp$^2$ structures between nanodiamond crystals [**8**], which can also be treated as a quasi-SL structure consisting of a non-periodic array of conducting (sp$^2$) layers. Some details of our proposed model can be found in our other report [**10**]. In this regard the microstructure is different to the conventional superlattice structure of compound semiconductors but similar to layered metals or superconductors which show strong anisotropic conductivities. Furthermore the density of planes increases with nitrogen concentration, which corresponds to an increase of coupling between planes that can be explained fairly well by a quasi-SL model [**17,18**]. We nevertheless point out that the disorder strength can be modified with the change of periodicity. In addition, we believe the highly conducting NCD samples belong to a family of diffusive (or weakly localized) transport regime and the small effect of activation can appear due to the energy difference between the nitrogen level and the conduction band (CB) (1$^{st}$ miniband) of the proposed quasi-SL (see Fig. 5, (b)) [**10**]. This concept has been verified from the MR study (see Fig**.** 4 and also in Ref. 10). A transition from VRH to WL has already been shown through the intentional introduction of disorder to artificial SL structures [**17,18**].

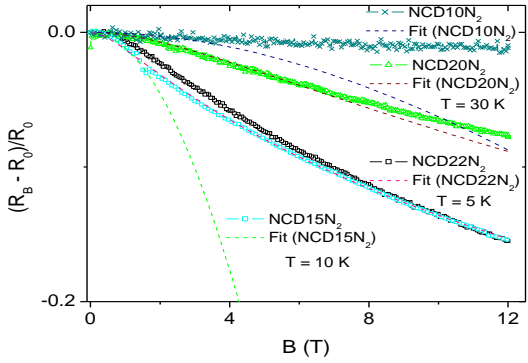

Fig.6: MR data for NCD samples fitted with the B-K model at different temperatures.

To identify the effect and quantify the degree of disorder in the UNCD samples we applied the Bryksin – Kleinert (B-K) model (Fig. 6) and analyzed the MR data for all samples [**19**]. At low temperatures the electronic transport mechanisms are dominated by elastic scattering from defects, impurities and electron-electron interactions. The B-K model showed in a self consistent theory that the ratio of the scattering times ($\frac{\tau_0}{\tau}$), where $\tau$ and $\tau_0$ correspond to the renomalized and elastic scattering times, respectively, would be a function of the anisotropy factor [**19**]. In that model (B-K) two guiding anisotropy parameters are used, i.e. $\alpha$, defined earlier and $\gamma = (1/\alpha)(k_z/k_{||})$ in which $k_z$, is the momentum cutoff parameter of the wave vector $k_z = 1/v_F\tau_e$ for any arbitrary field orientation [**19**]. Since the field was perpendicular to the samples only one anisotropy factor is used. Moreover, there is no restriction on the value of the disorder parameter, i.e. $q = k_F\lambda$. In this expression $\lambda$ and $k_F$ are the electron mean free path and Fermi wave vector respectively, which can vary from sample to sample, namely from NCD10N$_2$ to NCD22N$_2$ films [see Fig. 1(b)] and is actually a measure of order.

For small disorder i.e. $k_F\lambda \gg 1$ (i.e. in NCD22N$_2$ samples) the self consistency of the theory with respect to the renormalized scattering time $\tau$ might be disregarded yielding the corresponding conductivity expression through the Einstein relation $\sigma_{||} = e^2 D_{||} N(E_F)$ to be $\sigma_{||} = e^2 D_0 \frac{\tau}{\tau_0} N(E_F) = \sigma_0 \frac{\tau}{\tau_0}$, where $D_0 = \frac{1}{2}v^2\tau_0$. Hence the parallel component of conductivity is expressed as a function of the ratio $\frac{\tau}{\tau_0}$, implying a strong dependence on the anistropy in the material. The conductance pre-factor $G_0$ taken from the best fit of $G$ vs. $T$ data (using Eq. 1) was considered as the Drude contribution $\{\sigma_0 = e^2 D_0 N(E_F)\}$. For the metallic regime (e.g. NCD20N$_2$ and NCD22N$_2$ samples) $G_0$ was estimated on the basis of a constant term in the $G$ vs. $T$ fitting. For hopping transport expressed as $G(0,T) = G_{min} \exp(-\left(\frac{T_0}{T}\right)^{0.25})$, applicable to NCD10N$_2$ and NCD15N$_2$ samples, the estimation of $G_0$ was based on the $G_{min}$ prefactor, which had been considered as an analogue of $G_0$ in the metallic range. Further simplification was made by keeping the small disorder condition even in the hopping range and by replacing the renormalized scattering time $\tau$ with $\tau_0$. The results of the fit to the MR data for a set of NCD films prepared with different N$_2$ percentage based on the B-K model (given explicitly in Ref **19**) are shown in Fig 6. This model was found to work well at low fields. We took a constant value of $\alpha$ given by a PFS model fit for NCD22N$_2$ samples [see Eq. (3)], which is considered nearly the same for all samples. The value of disorder parameter $q \approx k_F\lambda$ is found to increase with conductivity of the NCD samples from 0.7 to 1.2, shown in inset of Fig. 1(c). A remarkable similarity between the conductance vs. N$_2$ concentration (measured at 56 K) curves and the $q$ vs. N$_2$ curve has been found, which clearly establishes the role of disorder in controlling transport in NCD films [Fig. 1(c)]. This analysis is found to be consistent with the Raman spectra of the film structures [Fig. 1(b)].

Furthermore the degree of disorder ($q = 0.95$) for NCD20N$_2$ was a little bit less than unity symbolizing the position of the $E_F$ in the localized states tails. Therefore, electrons require some activation energy to reach the extended states as found from the Arrhenius plots. For NCD10N$_2$ and NCD15N$_2$ samples, $q$ is found to be much less than unity, which means the disorder effect is large and hopping conduction can be dominant. The level of disorder found in NCD films from transport data analysis is consistent with the analysis of

Chimowa *et al.*

Raman spectra of the samples [Fig. 1(a) and (b)]. The disorder level in the films is found to decrease with the increase of $N_2$% used for the synthesis of the samples that increases the conductivity of the films. This difference of *q* suggests a possibility for a M-I transition in $NCD20N_2$ to $NCD10N_2$ samples. From the analysis of B-K model we revealed that the temperature dependence of $\tau_\phi$ is much weaker for the insulating regime of NCD films than for conducting one. The value of *D* is also found to decrease from 0.001 m$^2$/s to 0.0007 m$^2$/s as the samples change from metallic to hopping region. The elastic scattering time $\tau_0$ is of ~ 32 fs for all samples. Most importantly, $\tau_\phi$ increases by an order of magnitude in these samples at comparable value of temperature (0.028 ps at 10 K in $NCD15N_2$ to 0.5 ps at 5 K in $NCD22N_2$ samples). Such a long dephasing time and its weak temperature dependence show that these films are potential candidates for fast switching devices.

In conclusion, we have shown a conductivity crossover and related transport features in HFCVD grown NCD films which has not been seen so clearly in other NCD films. Within a large temperature range the electrical transport for all samples can be described by a disordered (non-periodic) quasi-SL model which can also explain the transport properties of amorphous carbon films [**20**] having a certain degree of disorder. The conduction for the $NCD20N_2$ films is typical of semiconductors where thermal activation predominates, while in the $NCD22N_2$ films we have semi-metallic behavior in which electron-electron interactions and 3D WL are corrections to the conductivity with the former dominating at low temperatures. We also observed a weak temperature dependence of dephasing time i.e., $\tau_\phi \sim T^{-0.7}$, which to our knowledge has been reported only in artificial superlattices [**18**]. Such behavior will give longer dephasing time in these films, a property which might help to realize these films in high-speed electronic devices. Furthermore, we have found the variation of disorder level derived from conductivity analysis in the samples is consistent with the Raman analysis. The low activation energy for the $NCD20N_2$ films is due to both the effect of disorder and the presence of shallow impurity band below the conduction band which is observed only in HFCVD films. This study in nanodiamond films lays a foundation for a class of novel fast switching devices and establishes the known conduction models even in HFCVD films, whose GB phase is different from that of MPCVD grown films.

\*\*\*

The authors are thankful to R.M. Erasmus for Raman spectroscopy and R. McIntosh for correcting the manuscript. S.B. thanks the NRF (SA) for granting the Nanotechnology Flagship Programme to perform this work and the University of the Witwatersrand Research Council for financial support.